\documentclass[%
preprint,
 amsmath,amssymb,
aps,
pra,
]{revtex4-1}

\usepackage[super]{nth}
\usepackage{textcomp}
\usepackage{graphicx}
\usepackage{epstopdf}

\begin{document}

\preprint{}

\title{Conduction through subsurface cracks in bulk topological insulators}

\author{S. Wolgast}
	\email{swolgast@umich.edu}
\author{Y. S. Eo}
\author{\c{C}. Kurdak}
\affiliation{University of Michigan, Dept.~of Physics, Ann Arbor, Michigan 48109-1040, USA}

\author{D.-J. Kim}
\author{Z. Fisk}
\affiliation{University of California at Irvine, Dept.~of Physics and Astronomy, Irvine, California 92697, USA}

\date{\today}

\maketitle

\section{\label{sec:Intro}Introduction}

Topological insulators (TIs) have the singular distinction of being electronic insulators while harboring metallic, conductive surfaces. In ordinary materials, defects such as cracks and deformations are barriers to electrical conduction, intuitively making the material more electrically resistive. Peculiarly, 3D TIs should become \emph{better} conductors when they are cracked because the cracks themselves, which act as conductive topological surfaces, provide additional paths for the electrical current. Significantly, for a TI material, \emph{any} surface or extended defect\cite{Ran} harbors such conduction. In this letter, we demonstrate that small subsurface cracks formed within the predicted\cite{Dzero2010,Takimoto,Lu,Fuhrman2014} 3D TI samarium hexaboride (SmB$_{6}$) via systematic scratching or sanding results in such an increase in the electrical conduction. SmB$_{6}$ is in a unique position among TIs to exhibit this effect because its single-crystals are thick enough to harbor cracks, and because it remarkably does not appear to suffer from conduction through bulk impurities.\cite{Wolgast,Kim13} Our results not only strengthen the building case\cite{Wolgast,Kim13,Zhang,Li,KimMag,Denlinger,Jiang,Neupane,XuSARPES,McQueen2015} for SmB$_{6}$'s topological nature, but are relevant to all TIs with cracks, including TI films with grain boundaries.

One of the biggest goals in the field of TIs has been to achieve materials in which the topologically-protected surface state is electrically accessible for device fabrication. However, most of the known 3D TIs suffer from finite bulk conduction, which pollutes or overwhelms the surface state in electrical measurements. Thus, a regime of surface dominated transport has only been achieved in ultra-thin films of these TIs. SmB$_{6}$ appears to be an exception with a truly insulating bulk and robust surface conduction.\cite{Wolgast,Kim13} Although this material has been studied as a mixed-valent insulator for more than four decades, it has been the subject of intense and renewed experimental study due to the recent theoretical\cite{Dzero2010,Takimoto,Lu,Fuhrman2014} and experimental\cite{Wolgast,Kim13,Zhang,Li,KimMag,Denlinger,Jiang,Neupane,XuSARPES,McQueen2015} evidence that it hosts a topologically protected surface state with a fully insulating bulk. Even though the topological nature of the surface state has been questioned,\cite{Rader} it is fair to say that the explosion of work on SmB$_{6}$ has established the TI hypothesis as providing the only unified framework in which to consider all the accumulated data for this material. Here we continue in this framework and exploit the truly insulating bulk behavior of SmB$_{6}$, coupled with the robust surface conduction on all of its surfaces, to explore some of the unusual properties of TIs with fully insulating bulk. We present not only the first direct evidence that there can be conduction through subsurface cracks, but also that such conduction can be the dominant transport.

Our results also have serious implications for researchers working on surface transport studies of SmB$_{6}$.\cite{Wolgast,Kim13,Thomas,Nakajima,Syers,Eo,WolgastMR,Yue,Greene} In fact, our efforts here were motivated by some of the puzzling results of such studies. For example, several transport studies, including our own,\cite{Wolgast} obtained carrier densities beyond the limit of $5.85\times10^{14}\text{ cm}^{-2}$ imposed on a single-band 2D system by the Brillouin zone size calculated for the lattice constant of SmB$_{6}$.\cite{Trounov} Reinterpretation of some historical transport studies\cite{Allen,Cooley95} in a 2D context also results in carrier densities which exceed this ``$k_{\text{F}}$'' limit.

This discrepancy is due in part to measurement difficulties unique to 3D TI bulk crystals, notably the possibility of conduction along unprepared side surfaces of a thinned sample with polished top and bottom surfaces. This problem was evaded using a Corbino disk geometry,\cite{WolgastMR,Eo} which is sensitive only to conduction along a single prepared surface. The polished surfaces on which the disks were fabricated yielded conductivities that were an order of magnitude smaller than indicated by a conventional Hall bar sample. While the corresponding carrier densities in our and others'\cite{Syers} Corbino disks were thereby rendered more physically reasonable, the new and surprising trend that we observed across our samples was that better polishing counterintuitively acted to diminish the surface conductivity. A similar trend has been observed on ion-damaged surfaces, for which the conductivity increases linearly with the depth of ion damage.\cite{LANLIons}

We hypothesized that the increase in conductivity with lower surface quality is due to cracks along the surface of the material. Surface roughening alone is not necessarily expected to change the 2D ``sheet'' carrier density (though it might decrease the carrier mobility), but any additional surfaces due to cracks along the visible top surface would contribute to the measured conductivity, as is the case with any 3D TI with no bulk conduction. Because such cracks would reside below the surface, they would not ordinarily be visible, but they would contribute to the total conduction as an apparent increase in the sheet carrier density and a corresponding reduction of the measured Hall coefficient.

\section{\label{sec:Exp}Scratch Experiment}

\begin{figure}[ht]
\includegraphics[width=8.5cm]{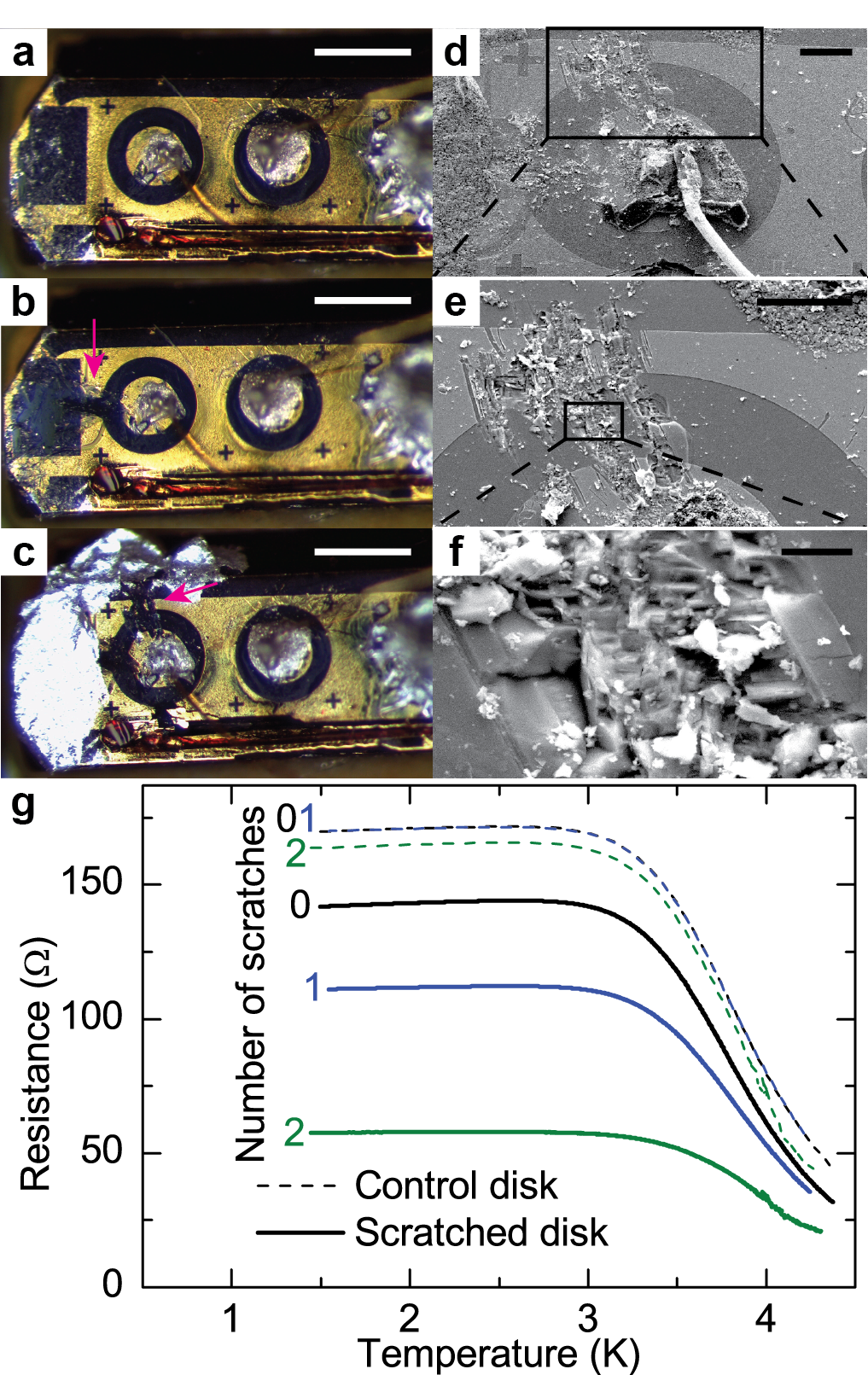}
\caption{Scratched Corbino measurements. \textbf{a} -- \textbf{c}, Optical images of the two Corbino disks before scratching and after each scratch. The scratches are indicated with magenta arrows. White scale bars are 500 \textmu m. \textbf{d} -- \textbf{f}, SEM images of the second scratch at increasing magnification. Black scale bars are 100 \textmu m, 100 \textmu m, and 10 \textmu m, respectively. \textbf{g}, Resistance vs temperature curves of both disks before scratching and after each scratch. Thick solid lines indicate the scratched disk, while thin dotted lines indicate the unscratched (control) disk.}
\label{fig:Scratch}
\end{figure}

In SmB$_{6}$ the surface conduction is manifested in its famous resistivity plateau that sets in below $3\text{ K}$. To verify the hypothesis of increased conductivity due to cracks near the surface, we studied the plateau over a series of resistance measurements on two Corbino disks fabricated on a single surface of SmB$_{6}$, shown in Fig.~\ref{fig:Scratch}a. After the temperature dependence of the resistances was obtained, the Corbino disks were brought to room temperature, and one was scratched between the inner and outer contacts in a radial direction using a diamond-tip scriber (Fig.~\ref{fig:Scratch}b). The disks were cooled and measured again, warmed and scratched a second time (Fig.~\ref{fig:Scratch}c), then cooled and measured a final time. Scanning Electron Microscope (SEM) images taken after the second scratch are shown in Fig.~\ref{fig:Scratch}d -- f. Resistance curves obtained after each scratch are plotted together for both Corbino disks in Fig.~\ref{fig:Scratch}g.

The plateau below 3 K, which corresponds to the resistance of the surface, decreases substantially after each successive scratch. Meanwhile, the second unscratched Corbino disk nearby on the same surface exhibits no change in each successive measurement. The resistance decrease in the first Corbino disk is opposite the change expected due to the accidental removal of portions of the gold contact pad by the scriber, but is consistent with the hypothesis that additional surface cracks could contribute to the total conduction. The magnitude of the change is also much larger than what is expected due to surface roughening, given that the scratched area is only a very small portion of the entire disk; such a large change suggests the presence of additional unseen conduction paths below the visible surface.

\begin{figure}[ht]
\includegraphics[width=8.5cm]{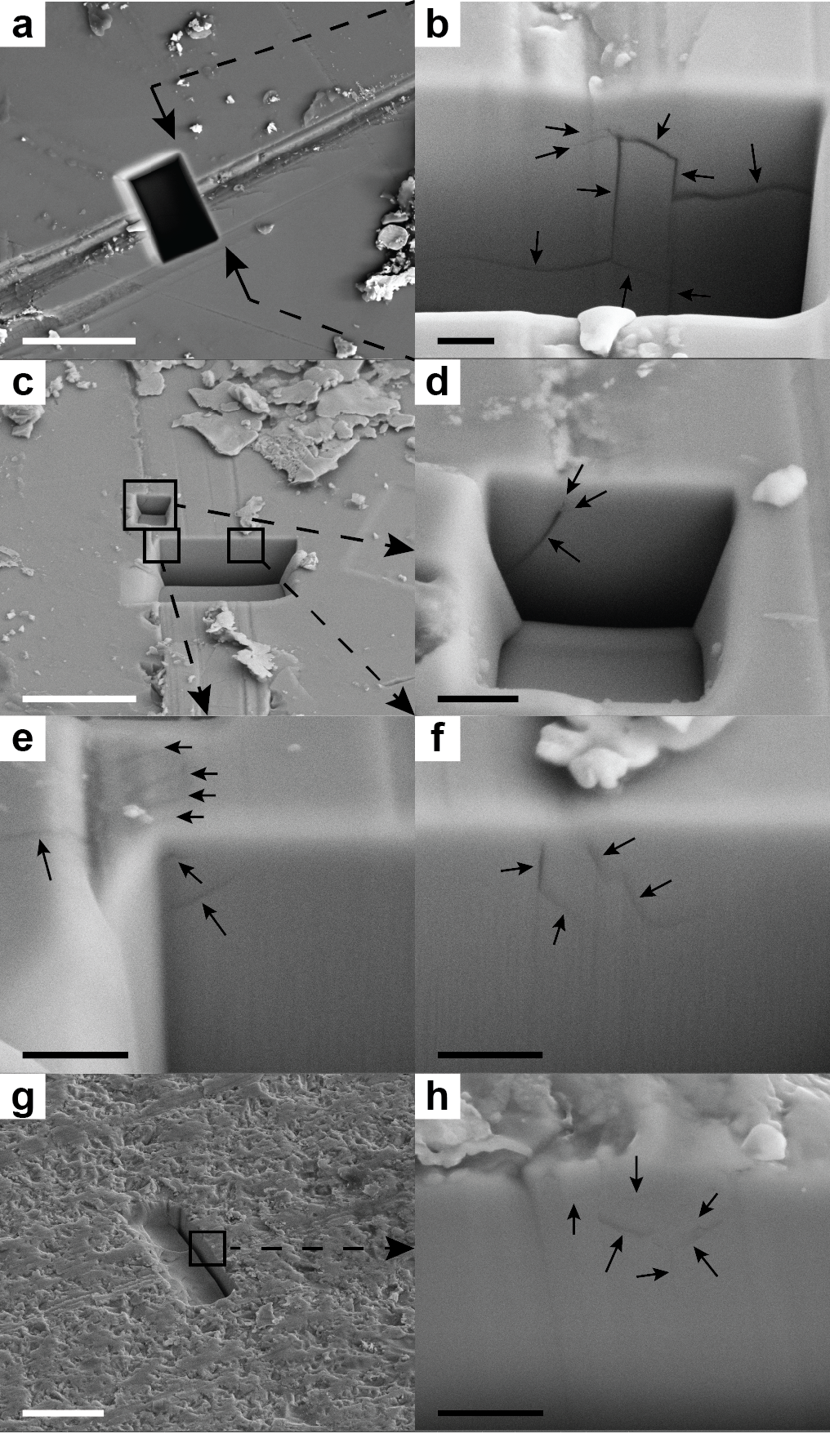}
\caption{Ion-milled cross-sections. \textbf{a}, Ion-milled cross-section across a scratch. \textbf{b}, Subsurface cracks visible below the scratch in \textbf{a}. \textbf{c}, Ion-milled cross-sections across another scratch. \textbf{d} - \textbf{f}, Subsurface cracks visible below the scratch in \textbf{c}. \textbf{g}, Ion-milled cross-section on a rough-polished surface. \textbf{h}, Subsurface cracks visible below the rough surface in \textbf{g}. White scale bars (panels \textbf{a}, \textbf{c}, and \textbf{g}) correspond to 10 \textmu m. Black scale bars (panels \textbf{b}, \textbf{d} -- \textbf{f}, and \textbf{h}) correspond to 1 \textmu m. In all panels, small black arrows indicate cracks.}
\label{fig:Cracks}
\end{figure}

Motivated by this result, we performed ion-beam milling on the scratched Corbino disk surface to reveal cross-sections of the scratch profile . Indeed, subsurface cracks are apparent below portions of the disk that were scratched, as shown in Fig.~\ref{fig:Cracks}a -- f. For example, the groove shown in Fig.~\ref{fig:Cracks}a is about 3-\textmu m wide and less than 0.5-\textmu m deep, but harbors subsurface cracks several \textmu m long in the transverse direction and up to 100-nm wide. Cracks shown in Fig.~\ref{fig:Cracks}e are visible both on and below the surface. Because the bulk transport behavior is unchanged by the scratches, and because the increased surface area due to surface roughening is not sufficient to explain the magnitude of the change in resistance we observe, we conclude that these cracks must be responsible for the additional conduction. Whether these cracks are oxidized when exposed to ambient air, as occurs for the outer surfaces,\cite{QMIXray,Phelan} is presently unknown; nevertheless, we expect that the transport properties of the cracks can be very different from those of the outer surface.

Subsurface cracks were also observed on a separate SmB$_{6}$ crystal prepared by rough-polishing (P1200 grit, which produces $\sim1$-\textmu m surface roughness), as shown in Fig.~\ref{fig:Cracks}g -- h. The cracks visible in Fig.~\ref{fig:Cracks}h are up to 1-\textmu m long in the transverse or vertical direction, though they are much narrower than those seen in Fig.~\ref{fig:Cracks}b and d, approaching the focal resolution limit of our SEM. Although cracks produced at finer polishing levels are below the resolution of our SEM images, it is not unreasonable to hypothesize their existence and contribution to the total surface conduction. Such cracks are expected to form whenever the maximum contact stress, which actually does occur a few \textmu m below the surface,\cite{Hertz} exceeds the tensile strength of the material. 

\section{\label{sec:Conc}Conclusions}

The most important implication of this observation is that any transport study on surface conductors such as 3D topological insulators with an insulating bulk must account for the possibility of cracks such as these, especially since such cracks may have different transport parameter values than the outer surface. In particular, this result provides a way to account for the reduced conductivity in SmB$_{6}$ as surface quality is improved, as well as the extraordinarily high apparent conductivity and unphysically large carrier density (small Hall coefficient) in our Hall bar measurements. This situation might be exacerbated if such cracks make electrical contact with hidden aluminum inclusions in flux-grown samples that would otherwise be insulated from the surface transport. However, the complications presented by the possibility of cracks extends also to crystals grown via the floating zone melting process. 

We have demonstrated that subsurface cracks occur near the surface of SmB$_{6}$, and have given strong arguments based on experiment that these cracks can greatly enhance the surface conduction in electrical transport measurements. Specifically, surface damage due to scratches or other roughening mechanisms produce cracks that act as additional conductive pathways for intrinsic surface transport materials such as topological insulators, and this additional conduction results in a higher apparent carrier density in transport measurements.

\section{\label{sec:Methods}Methods}

The crystals used in this study were grown via the aluminum flux method. The (011) crystallographic surface of the single-crystal was prepared by polishing with silicon carbide abrasive pads in incrementally smaller grit sizes to P4000, and two Corbino disks with inner diameters of 300 \textmu m and outer diameters of 500 \textmu m were fabricated side-by-side on the polished surface using standard photolithography. 50/1500 {\AA} Ti/Au contacts were deposited by evaporation to form the metalized portion of the disks, and they were wirebonded by hand using silver paint and silver epoxy. The Corbino disks' resistances were measured in a variable-temperature cryostat between 1.5 K and 4 K using standard lock-in techniques at a frequency of 17.72 Hz. Between each measurement, the surface was scratched by hand with a diamond scriber. The additional silver paint surrounding the disk in Fig.~\ref{fig:Scratch}c was added to ensure electrical continuity with the outer Ti/Au contact pad. Ion-beam milling and subsequent electron imaging was performed on an FEI Nova 200 Nanolab SEM/FIB.

\bibliography{CrackReferences}

\section*{\label{sec:Acknowledgements}Acknowledgements}

The authors thank Gang Li and Fan Yu for the optical imaging, J. W. Allen for helpful discussion and review of the manuscript, and Leonard Sander for his insight on crack formation. The samples were prepared in part at the Lurie Nanofabrication Facility, a member of the National Nanotechnology Infrastructure Network, which is supported by the NSF. The ion beam milling and SEM images were obtained at the Electron Microbeam Analysis Laboratory, supported in part by NSF grant \#DMR-0320740. Funding was provided by NSF grants \#DMR-1006500, \#DMR-1441965, and \#DMR-0801253.

\end{document}